# Vortex Phase Diagram of Layered Superconductor $Cu_{0.03}TaS_2$ for $H \parallel c$

X. D. Zhu[a,*], J. C. Lu[b], Y. P. Sun[a,c], L. Pi[b], Z. Qu[a], L. S. Ling[a], Z. R. Yang[c], Y. H. Zhang[a,b,*]

[a] High Magnetic Field Laboratory, Chinese Academy of Sciences, Hefei 230031, People's Republic of China

[b] Hefei National Laboratory for Physical Sciences at Microscale, University of Science and Technology of China, Hefei 230026, People's Republic of China

[c] Key Laboratory of Materials Physics, Institute of Solid State Physics, Chinese Academy of Sciences, Hefei 230031, People's Republic of China

*Corresponding author at: High Magnetic Field Laboratory, Chinese Academy of Sciences, 230031, P. R. China. Tel.: +86 551 559 5619; fax: +86 551 559 1149.

Email: xdzhu@hmfl.ac.cn; zhangyh@ustc.edu.cn;

[Abstract] The magnetization and anisotropic electrical transport properties have been measured in high quality $Cu_{0.03}TaS_2$ single crystal. A pronounced peak effect has been observed, indicating that the high quality and homogeneity are vital to peak effect. A kink has been observed in the magnetic field $H$ dependence of the in-plane resistivity $\rho_{ab}$ for $H \parallel c$, which corresponds to a transition from activated to diffusive behavior of vortex liquid phase. In the diffusive regime of the vortex liquid phase, the in-plane resistivity $\rho_{ab}$ shows $\rho_{ab} \propto H^{0.3}$ relation, which does not follow the Bardeen-Stephen law for free flux flow. Finally, a simplified vortex phase diagram of $Cu_{0.03}TaS_2$ for $H \parallel c$ is given.

[Introduction]

When a type-II superconductor is placed under a magnetic field $H$ above the lower critical field ($H_{C1}$) and below the upper critical field ($H_{C2}$), the magnetic field penetrates into the superconductor through the vortex arrays, each of which carries one quantum flux surrounded by circulating supercurrent [1]. This state is called the *mixed state*. Now, it is well known that for many type-II superconductors, the *mixed state* is actually composed of many complex vortex phases, such as vortex solid phase, vortex glass phase and vortex liquid phase [2], instead of the simple vortex lattice predicted by Abrikosov [3]. Although these phases have been studied extensively in the past decades, the details and the transitions between these phases remain issues of debate.

Peak effect (PE), referring to the anomalous increase and the pronounced maximum of critical current density ($J_C$) or magnetization prior to the irreversible field ($H_{irr}$), has been observed in many type-II superconductors, such as Nb [4, 5], CeRu$_2$ [6, 7], V$_3$Si [8], layered NbSe$_2$ [9-13], $Re$Ni$_2$B$_2$C ($Re$ = Dy, Ho, Er, Tm, Y, Lu) [14, 15], MgB$_2$[16]. In addition, similar phenomena have been observed in high temperature cuprate superconductors (HTCS) [17-23], and the recent discovered iron based superconductors [24, 25] *etc.*, which are believed to have different origins and is usually called as *fishtail effect*. Experimentally, PE for layered superconductors is usually observed in weakly pinned and high quality single crystals for $H$ being along the $c$-axis ($H \parallel c$). An incredible amount of publications dealing with the underlying physics of PE have been published during past decades, but the interpretation remains a controversial issue [26-33]. So far, it is widely accepted that the PE is related to a vortex phase transition, though the details of the vortex phases remain controversial. Among the proposed mechanisms of PE, phenomenological pictures based on an order-disorder (OD) transition from a quasi-ordered Bragg glass (weakly pinned elastic glass) phase [30] to a disordered phase with proliferation of topological defects [31,32] (or with these two co-existing phases [33]) explain a broad number of related experimental results.

The vortex liquid phase lies between $H_{irr}$ and $H_{C2}$. Phase transition not only occurs between vortex solid (or vortex glass) and vortex liquid phases, but also between different regimes of vortex liquid phases. An unusual reversible second order phase transition between two vortex liquid phases has been discovered in $YBa_2Cu_3O_7$ single crystal by heat capacity and magnetization measurements [34]. In addition, the transition between two different regimes of vortex liquid phase has been observed from the magnetic field dependence of resistivity ($\rho$) behavior and the abnormal negative minimum of Hall resistivity in $YBa_2Cu_3O_7$. It was explained in a phenomenological picture of transition from "activated" to "diffusive" behavior of vortex motion [35]. However, because of the extremely large $H_{C2}$, the limitation of the applied magnetic field and the fluctuation of superconductivity in HTCS, normalized superconducting transition temperature $t$ ($T/T_C$, where $T$ is the temperature, $T_C$ is the superconducting transition temperature) is rather limited in $YBa_2Cu_3O_7$ ($t > 0.84$) [35]. Hence, the $\rho$-$H$ relation for lower $t$ remains unexplored. Interestingly, a similar transition in 2H-NbSe$_2$ single crystal was observed from the magnetoresistance results for $H \parallel c$ (measured at $T = 4.2$ K) [36], though it was not discussed.

In our previous work, high-quality $Cu_{0.03}TaS_2$ single crystals with $T_C^{onset} = 4.2\,\text{K}$ were grown successfully [37], which is isostructural with 2H-NbSe$_2$. The Ginzburg number ($G_i$) is a measure of the importance of thermal fluctuations according to the traditional collective pinning theory [27], and is given by the formula $G_i = [k_B T_C / H_C^2(0)\xi_c(0)\xi_{ab}^2(0)]^2 / 2$. The estimated $G_i$ for $Cu_{0.03}TaS_2$ is of the order of $10^{-5}$ [37], comparable with that for $NbSe_2$ (~$3\times10^{-4}$) [6]. Thus, $Cu_{0.03}TaS_2$ should be a good candidate for PE.

When the Lorentz-Magnus force ($F_L = J \times B$, where $B$ is the magnetic flux density) exerted by the current density ($J$) overcomes the pinning forces, the vortices will move viscously with a mean velocity $V_L$ generating an electric field $E = B \times V_L$ [38]. The viscous friction depends on the energy dissipation process of quasi-particles in and around the normal core of the vortices. Therefore, the dc resistivity $\rho = E/J$ measured in the vortex liquid phase is related to the magnetic field and the vortex motion (or flux flow). The flux flow resistivity ($\rho_f$) is defined as the slope of the $E-J$ curve ($\partial E/\partial J$) to manifest the resistivity which is independent of the pinning force [38]. Bardeen-Stephen (BS) law predicts that if other forces such as vortex-defect interaction (pinning) are negligible, the flux will flow freely, and the free flux flow (FFF) resistivity will be proportional to the magnetic field $H$, $\rho_f = \rho_n H/H_{C2}$, where $\rho_n$ is the normal state resistivity [39]. The BS law has been confirmed in most dirty and clean s-wave systems. In an extremely clean $NbSe_2$ single crystal, the confirmation of BS law has been observed from the magnetic field dependence of resistivity and Hall effect [40]. However, in some layered superconductors with gap nodes or with multiple bands gap, such as $Bi_2Sr_2CuO_{6+\delta}$ [41], $Bi_2Sr_2CaCuO_{8+\delta}$ [42], and $MgB_2$ [43], the BS law breaks down in the micro-wave impendence experiments. The origin of these phenomena has been attributed to the different energy dissipation influenced by the abnormal gap structures [41].

In this study, we report the discovery of PE and the transition from activated to diffusive behavior of vortex liquid phase in $Cu_{0.03}TaS_2$ for $H \parallel c$. Finally, a simplified vortex phase diagram of $Cu_{0.03}TaS_2$ is proposed.

[Experimental] Platelets of single crystal $Cu_{0.03}TaS_2$ with $T_C$ = 4.2 K used in this study were grown via iodine vapor transport method as described in our previous report [37]. The onset superconducting transition temperature is 4.2 K with a transition width (10%-90%) of ~ 0.1 K, indicating the high quality of the sample. The measurements discussed in this study were carried out on a sample with a dimension of 3.18 mm (*l*, the longest dimension) × 1.25 mm (*w*, the width) × 0.34 mm (*t*, the shortest dimension) mm$^3$ with *t* along the *c*-axis. The isothermal dc magnetization hysteresis measurements were performed using a Quantum Design Superconducting Quantum Interference Device (SQUID) system (1.8 K ≤ $T$ ≤ 300 K, 0 T ≤ $H$ ≤ 7 T). The $H$ dependence of the $J_C$ were extracted from the isothermal magnetization loop results using the formula $J_C = 20(M^+ - M^-)/w(1-\frac{w}{3l})$ according to the Bean critical state model [44], where $M^+$ and $M^-$ represent the magnetization (emu/cm$^3$) measured during the process of decreasing and increasing field, respectively, and *w* and *l* measured in cm.

The anisotropic transport properties measurements were performed using standard four probe method in the Quantum Design Physical Property Measurement System (PPMS) (1.8K ≤ $T$ ≤ 300K, 0T ≤ $H$ ≤ 16T). In order to make sure that the direction of $H$ would be exactly parallel to the *ab* plane (*H*||*ab*) and the *c*-axis (*H* || *c*) of the single crystal, the rotating sample holder was used. We measured the angular dependence of the in-plane resistivity ($\rho_{ab}$) at $T$ = 3.7K and $H$ = 0.8 T to determine the direction of the magnetic field. As shown in Figure 1, *H*||*ab* was determined as the angles corresponding to the minimal resistance ($\theta_{ab}$), and $H$ || $c$ was determined as $\theta_{ab}$ plus 90 degrees. During all the measurements, the excitation current was kept at 5 mA ($J \approx 1.1$A/cm$^2$), and the contact resistance was less than 1Ω.

[Results and Discussions]

*1. Peak effect*

Figure 2 shows the isothermal superconducting magnetization hysteresis (*M-H*) loop for $Cu_{0.03}TaS_2$ measured at *T* = 2.0 K. As marked by an ellipse in Figure 2, a pronounced anomalous PE feature is observed. The high reversibility around the PE region indicates the high quality of the sample. The right inset panel shows *M-H* plot on an expanded scale to emphasize the presence of PE anomaly around a field of 1.1 T. The fields corresponding to the onset and peak of the PE are marked as $H_P^{onset}$ and $H_P$, respectively. $H_{irr}$ is estimated from the field where the two branches of the hysteresis loops meet, as shown in the right inset of Figure 2. The obtained $H_P^{onset}$, $H_P$, $H_{irr}$ are ~ 0.98 T, 1.1 T, 1.3 T, respectively.

In order to investigate the PE further, more *M-H* loops were measured at different temperatures. Figure 3 shows the portion of *M-H* loops (the first and fourth quadrants) for $Cu_{0.03}TaS_2$ measured from *T* = 2.0 to *T* = 3.6 K. All the curves are shifted for clarity except for *T* = 2 K. The normalized data of $j_C$ ($J_C / J_C^{h=0}$, where $J_C^{h=0}$ represents the critical current density under zero field) as a function of normalized field ($h = H/H_{irr}$) are extracted from the $J_C(T, H)$ data.

Figure 4 shows log-log plots of the *h* dependence of $j_C$ at different temperatures for $Cu_{0.03}TaS_2$. Obviously, in the range of 0.01 < *h* < 0.1, the $j_C$–*h* relation follows a power law relation $j_C \propto h^{-n}$ (with *n* ≈ 1), and overlaps very well except the PE region for different temperatures. Interestingly, the power law behavior of $j_C$–*h* has been discovered in $NbSe_2$ [45] and $SnMo_6S_8$ [46], in agreement with predictions of the weak collective pinning theory [27] that attributes its origin due to the inter vortex interactions [47]. Individual pinning should be dominant below the weak collective pinning region, which is separated by a kink on the $J_C$-$H$ curve [46]. In the PE region, the value of the normalized peak position ($H_P/H_{irr}$) of PE is almost unchanged ($h \sim 0.71$) initially, but it decreases gradually with increasing temperature.

The PE has not been observed in directly synthesized $Na_xTaS_2$ [48], or $Ni_xTaS_2$ grown from NaCl/KCl flux [49], whose superconducting transition widths are larger than that of $Cu_{0.03}TaS_2$ [37]. It indicates that the high quality and homogeneity of the sample are necessary for PE. The inhomogeneity of intercalates leads to the $T_C$ fluctuation in real space and provides additional pinning centers when the $T$ approaches $T_C$. The high density inhomogeneity caused by intercalates will induce disorder, and lead to the disappearance of the OD transition. It will further lead to the disappearance of PE near $H_{irr}$. This can also explain why only 200 ppm of Fe doping causes a significant effect of PE broadening and weakening in $NbSe_2$ [50]. Thus, our data support that OD transition is the origin of the PE in $Cu_{0.03}TaS_2$.

*2. The transition from activated to diffusive behavior of vortex liquid phase*

Figure 5(a) depicts the *M-H* curve and the magnetic field dependence of $\rho_{ab}$ measured at T = 2.0 K. The inset is the sketch of the sample arrangement with respect to the *H* and its contacts. Figure 5(b) depicts the log-log plot of $\rho_{ab}$-*H* curve, whose inset shows the determination of $H_{C2}$. As shown in the Figure 5(a), with increasing *H*, the $\rho_{ab}$-*H* curve can be divided into three regimes: the first one is the superconducting regime below $H_{irr}$, where $\rho_{ab}$ remains zero; in the second one, $\rho_{ab}$ starts to increase abruptly from $H_{irr}$ = 1.3 T to a kink at $H \sim$ 1.9 T; in the third one, $\rho_{ab}$ increases slowly with a power law up to $H_{C2}$ = 6.95 T (as shown in Figure 5(b)).

Figure 6(a) and Figure 6(b) show the measured $\rho_{ab}$–*H* curves at different temperatures for *H*||*ab* and *H* || *c*, respectively. The insets of Figure 6(a) and Figure 6(b) show the sketches of the

direction of the $H$, $J$ and the vortex motion. As shown in the Figure 6, with increasing $T$, the superconducting transition moves to lower field both for $H \parallel ab$ and $H \parallel c$. However, there is no kink in the flux flow region of the $\rho_{ab}$–$H$ curve for $H \parallel ab$.

Apparently, the kink only occurs in the vortex liquid phase for $H \parallel c$, which should be related to a vortex phase transition. We define the $H$ corresponding to the kink as $H_k$. Interestingly, no anomaly can be observed from the $M$-$H$ curves at $H_k$. In pure and high quality $Bi_2Sr_2CaCu_2O_8$ [51] and $YBa_2Cu_3O_7$ [52] single crystals, there is a transition in the temperature dependence of resistivity under magnetic field, separating the abrupt increasing region from zero resistivity and the slowly broadening region. The transition has been reported to originate from the first order melting of the vortex lattice, with a characteristic of a discontinuous change of the magnetization [51]. However, as shown in Figure 5, no such discontinuous change of the magnetization can be observed from the $M$-$H$ curve for $Cu_{0.03}TaS_2$. Thus, the transition from the vortex liquid to another vortex liquid phase observed in $Cu_{0.03}TaS_2$ should be attributed to the transition of activated to diffusive behavior in the vortex liquid phase rather than the first order melting of the vortex lattice.

In the activated regime, the vortex line is activated from the strong pinning barrier, which leads to the abrupt increase of $\rho_{ab}$ from zero with increasing $H \parallel c$. The analysis the activated regime of $\rho_{ab}$-$H$ curve with the thermal activated plastic motion model did not give any satisfactory fits [53-55]. Therefore, the temperature dependence of $\rho_{ab}$ is measured to obtain the thermal activation energy ($U_{act}$). The activated regime in the vortex liquid phase is more obvious in $\rho_{ab}$–$T$ curves than those in $\rho_{ab}$–$H$ curves. Figure 7 depicts the $H$ dependence of the obtained $U_{act}$, which is obtained from the $ln\rho_{ab}$ – $1/T$ curves shown in the inset according to $\rho \approx \rho_n e^{-U_{act}/T}$ [53].

The $H$ dependence of $U_{act}$ fits well with the power law $U_{act} \sim H^{1.04}$. The $U_{act}$ of Cu$_{0.03}$TaS$_2$ has a typical value of 300-900 K, which is far from the experimental temperature. The plastic barriers $U_{pl}$ and $T_m$ (the first order thermal melting temperature) can be related via Lindemann number [2, 56]: $T_m \cong 2.7 c_L^2 U_{pl}$. Assuming $U_{pl}$ is associated with $U_{act}$, and $c_L \sim 0.2$, an unreasonable value of 30 - 90 K is derived for $T_m$, which further confirms that the transition in the vortex liquid phase does not originate from the first order thermal melting.

In the diffusive regime, the thermal activated plastic motion model is not valid, and the $\rho_{ab}$ mainly depends on the energy dissipation of vortex motion for $H \parallel c$, indicating a broad transition. In contrast, for $H \parallel ab$, crossing the Van der Waals gap between the superconducting TaS$_2$ layers provides the main energy dissipation because of the strong intrinsic pinning. Thus, for $H \parallel ab$, the transition of activated to diffusive behavior in the vortex liquid phase does not appear in both 2H-NbSe$_2$ [36] and Cu$_{0.03}$TaS$_2$.

*3. The breakdown of the Bardeen-Stephen Law*

Figure 8 shows the $h$ dependence of the normalized resistivity $\rho_{ab}/\rho_n$ ($\rho_n$ represents the $\rho_{ab}$ at $T$ = 5 K) with $H \parallel c$. As shown in the Figure 8, in the activated regime, $\rho_{ab}/\rho_n$–$h$ curves measured at different temperatures overlap with each other; in the diffusive regime, the curves almost overlap with each other when $T \leq 3.2$K, and then turns up rapidly when $T > 3.2$ K. This can be explained by the extra energy dissipation that contributes to the flux flow resistivity when $T$ is near $T_C$. The inset shows the log-log plot of the $h$ dependence of the normalized resistivity $\rho_{ab}/\rho_n$ for $H \parallel c$. Obviously, $\rho_{ab}/\rho_n$-$h$ curves of the diffusive regime are almost logarithmic when $T \leq 3.2$K. The $\rho_{ab}/\rho_n$-$h$ curves can be fitted to $f(h) = h^{0.3}$ very well when $T < 3$ K, which is depicted as a dashed curve in Figure 8. Although lower temperature cannot be achieved due to the instrumental limit, our results strongly suggest that $\rho_{ab}/\rho_n$-$h$ relation for $T\to 0$ should be also logarithmic and near $h^{0.3}$. It should be noted that, the $\rho_{ab}$-$H$ curve of YBa$_2$Cu$_3$O$_7$ at $t \sim 0.85$ is very similar to that of Cu$_{0.03}$TaS$_2$ for $t > 0.8$ ($T > 3.2$K) [35]. For YBa$_2$Cu$_3$O$_7$, the fluctuation will be weaker and the vortex phases will be far from the multi-critical point of the phase diagram at lower $t$ [2]. But the $\rho_{ab}$-$H$ of YBa$_2$Cu$_3$O$_7$ may have different relation. Thus, more experiments with higher magnetic field and lower $t$ are needed to study the vortex liquid phase for YBa$_2$Cu$_3$O$_7$ and other HTCS.

As discussed above, in the diffusive regime of the vortex liquid phase, the motion is almost free, and the $\rho_{ab} \propto \rho_n(H/H_{C2})^{0.03}$ behavior in Cu$_{0.03}$TaS$_2$ does not follow the BS law for free flux flow. Assuming that in the diffusive regime only flux flow contributes to the $\rho_{ab}$ and $B$ equals to $H$, $V_L \propto H^{-2/3}$ relation can be deduced from $\rho_{ab} = E/J = B \times V_L / J$. Thus, with increasing $H$, $V_L$ increases in the activated regime from $H_{irr}$, then reaches a peak at $H_k$, and further decreases with an relation of $V_L \propto H^{-2/3}$.

Interestingly, the Na$_x$TaS$_2$ and the Ni$_x$TaS$_2$ single crystal do not show any sign of the vortex liquid to liquid transition. Therefore, the inhomogeneity makes vortex-defect interaction the primary energy dissipation channel as opposed to the flux flow, and therefore the diffusive regime is not established. Thus, it can be concluded that, the high quality and weaker pinning of the sample is vital to the PE and the transition from activated to diffusive behavior of vortex liquid phase.

*4. The vortex phase diagram*

Figure 9 shows the vortex phase diagram for Cu$_{0.03}$TaS$_2$, depicting the temperature dependence of $H_p^{onset}$, $H_p$, $H_{irr}$, and $H_{C2}$ [57]. All critical fields have almost linear $T$ relations in the experimental temperature range. The vortex phases of the PE region are just named as the "Peak Effect". Based upon the discussions above, the vortex phase diagram is composed of the Bragg glass phase, the PE region, the activated and the diffusive regimes of vortex liquid phase.

2H-NbSe$_2$ shares the same structure with Cu$_{0.03}$TaS$_2$. So far, PE in 2H-NbSe$_2$ has been extensively studied, but not the vortex liquid phase. Our results suggest that the vortex liquid phase of 2H-NbSe$_2$ may also be complicate as in Cu$_{0.03}$TaS$_2$ and HTCS.

[Summary]

In summary, peak effect was observed from the superconducting magnetic hysteresis loops of Cu$_{0.03}$TaS$_2$. A transition from activated to diffusive behavior of vortex liquid phase was observed from the magnetoresistance experiment for $H \parallel c$. In the diffusive regime of the vortex liquid phase, the in-plane resistivity $\rho_{ab}$ shows $\rho_{ab} \propto H^{0.3}$ relation, which does not follow the Bardeen-Stephen law for free flux flow. Finally, a simplified vortex phase diagram of Cu$_{0.03}$TaS$_2$ for $H \parallel c$ is given. Our results indicates that high density disorder in TaS$_2$ intercalated superconductors will lead to the disappearance of PE and the transition from activated to diffusive behavior of vortex liquid phase.

[Acknowledgements]

This work was supported by the State Key Project of Fundamental Research of China (2007CB925001, 2010CB923403, and 2007CB925002), Director's Fund of Hefei Institutes of Physical Science, Chinese Academy of Sciences ("Spark foundation" No.Y06CS51131B$_2$Q), and the National Nature Science Foundation of China under contract No.10774146, 10774147.

[57] We neglect the Meissner phase and the vortex liquid phase just above the Meissner phase that is not investigated here.

Figure captions:

Figure 1. The angular dependence of the in-plane resistivity ($\rho_{ab}$) at $T$ = 3.7K and $H$ = 0.8T. The arrows mark the degrees corresponding to $H||ab$ and $H || c$. The insets show the sketches of the sample arrangement with respect to the direction to the field.

Figure 2. Superconducting *M-H* loop for $Cu_{0.03}TaS_2$ measured at $T$ = 2.0 K with $H//c$. The arrows on the curve show the processes of increasing field and decreasing field during the measurement. The left inset shows the dimensional sketch of sample with respect to the field. The PE region is marked by an ellipse. The right inset shows the magnified plot of the *M-H* in the vicinity the PE region. $H_P^{onset}$, $H_p$, $H_{irr}$ are marked by arrows.

Figure 3. The magnified plots of *M-H* loops for $Cu_{0.03}TaS_2$ measured at different temperatures. All curves are shifted for clarity, and the dashed lines represents $M$ = 0 at each temperature.

Figure 4. The log-log plot of the normalized critical current density ($j_C = J_C / J_C^{h=0}$, where $J_C^{h=0}$ represents the $J_C$ measured when $H$ = 0) versus normalized field $h$ ($h = H/H_{irr}$). The dashed line serves as a guide for the eye. The arrow shows the direction of temperature increasing.

Figure 5. (a) The *H* dependence of *M* (left panel) and $\rho_{ab}$ measured at 2.0 K with $H || c$. The $H_{irr}$ is shown by a dotted line, and the $H_k$ is marked by an arrow. (b) The log-log plot of the $\rho_{ab}$-*H* curve. The dashed line serves as a guide for the eye. The $H_{C2}$ is marked by an arrow. Inset shows the sketch of sample arrangement with respect to the *H* and its contacts.

Figure 6. The $H$ dependence of the activation energy $U_{act}$ for $Cu_{0.03}TaS_2$. The dashed line represents the fitting to $U_{act} \sim H^{-1.04}$. The inset shows the $ln\rho_{ab}$ versus $1/T$ for $Cu_{0.03}TaS_2$.

Figure 7. The $H$ dependence of $\rho_{ab}$ with $H||ab$(a) and $H || c$(b) measured at different temperatures. The insets in (a) and (b) shows the sketches of the sample arrangement with respect to the $H$, the probing current, and the moving direction of the vortices.

Figure 8. The normalized field $h = H/H_{C2}$ dependence of normalized in-plane resistivity ($\rho_{ab}/\rho_n$) measured at different temperatures. The dashed curve represents the function $f(h) = h^{0.3}$. Inset shows log-log plot of the $H$ dependence of $\rho_{ab}/\rho_n$ measured at different temperatures.

Figure 9. The $H$-$T$ vortex phase diagram for $Cu_{0.03}TaS_2$, depicting the temperature dependence of $H_p^{onset}$, $H_p$, $H_{irr}$, and $H_{C2}$. For details of the different vortex states, see the text.

Figures:

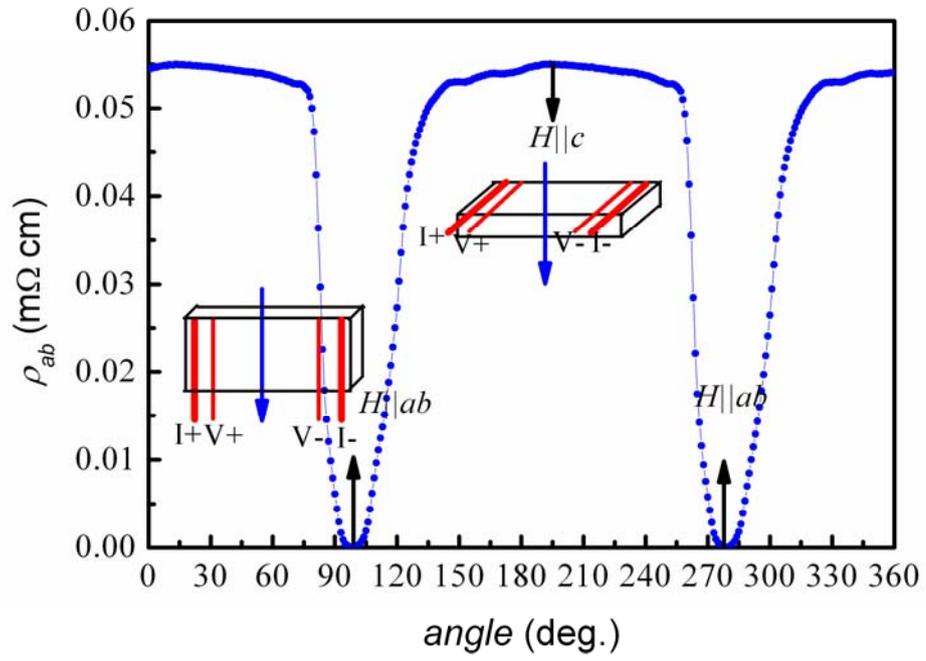

Figure 1 X. D. Zhu *et al.*

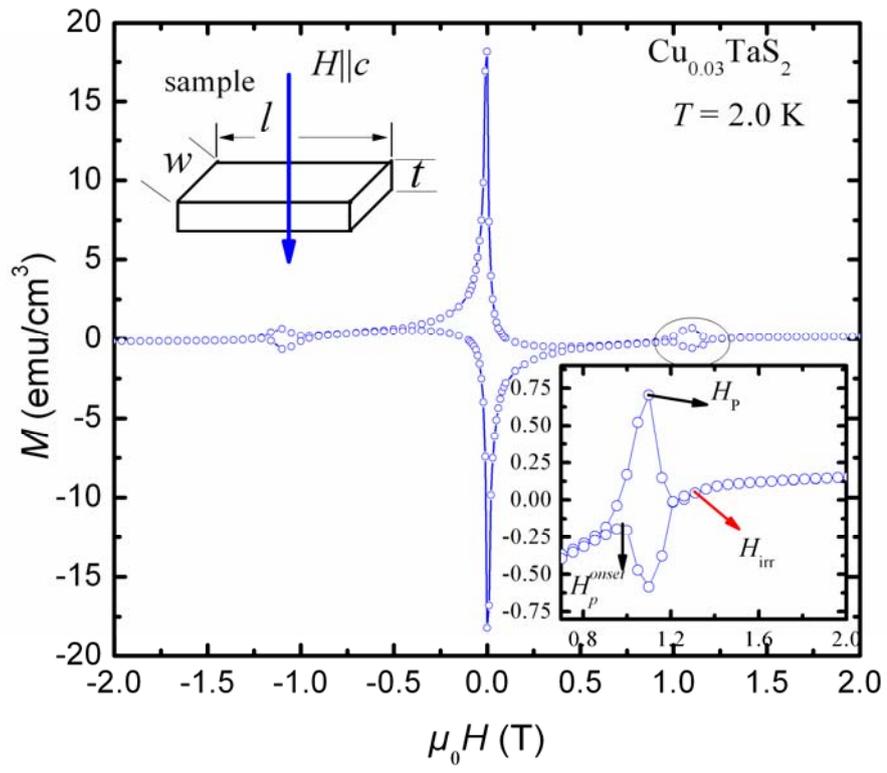

Figure 2 X. D. Zhu *et al.*

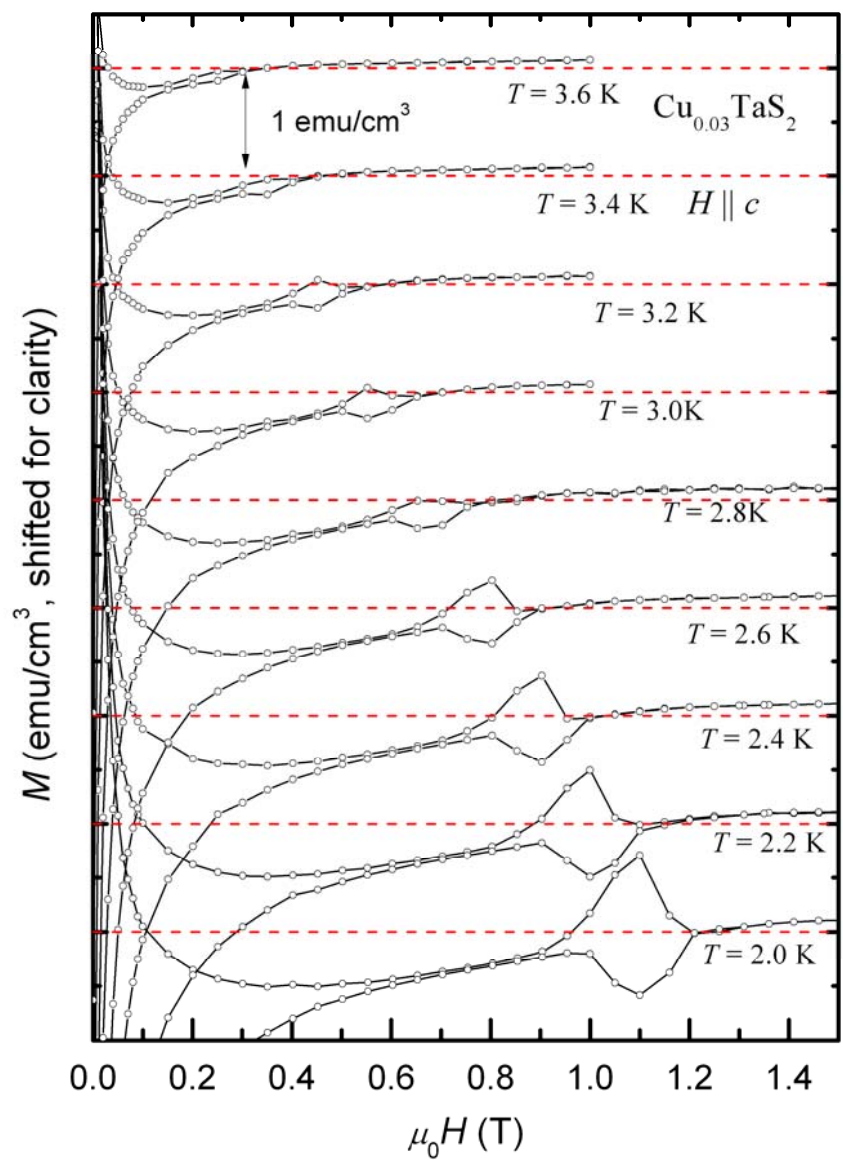

Figure 3 X. D. Zhu *et al.*

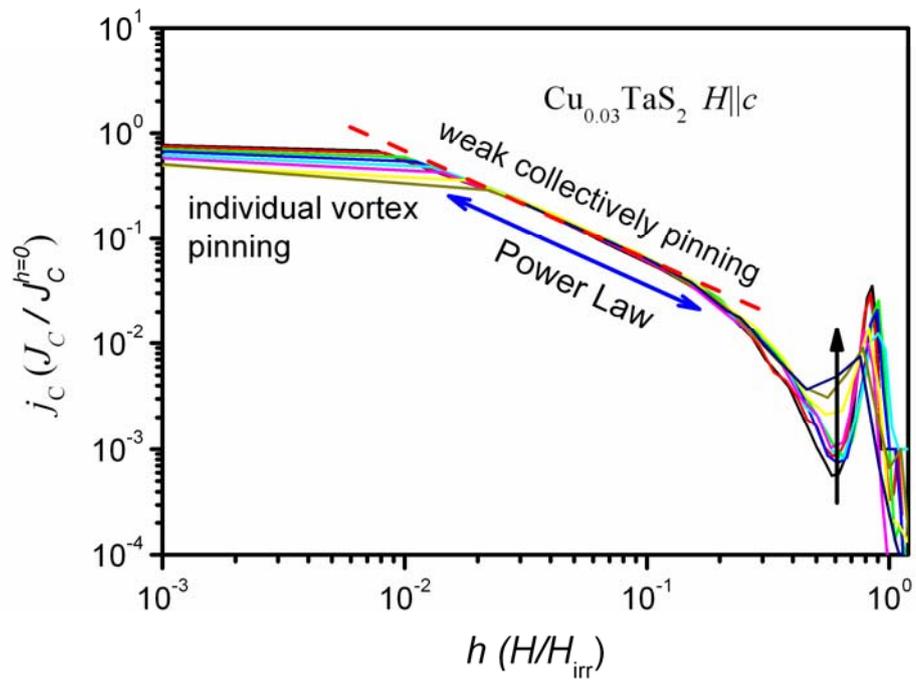

Figure 4 X. D. Zhu *et al.*

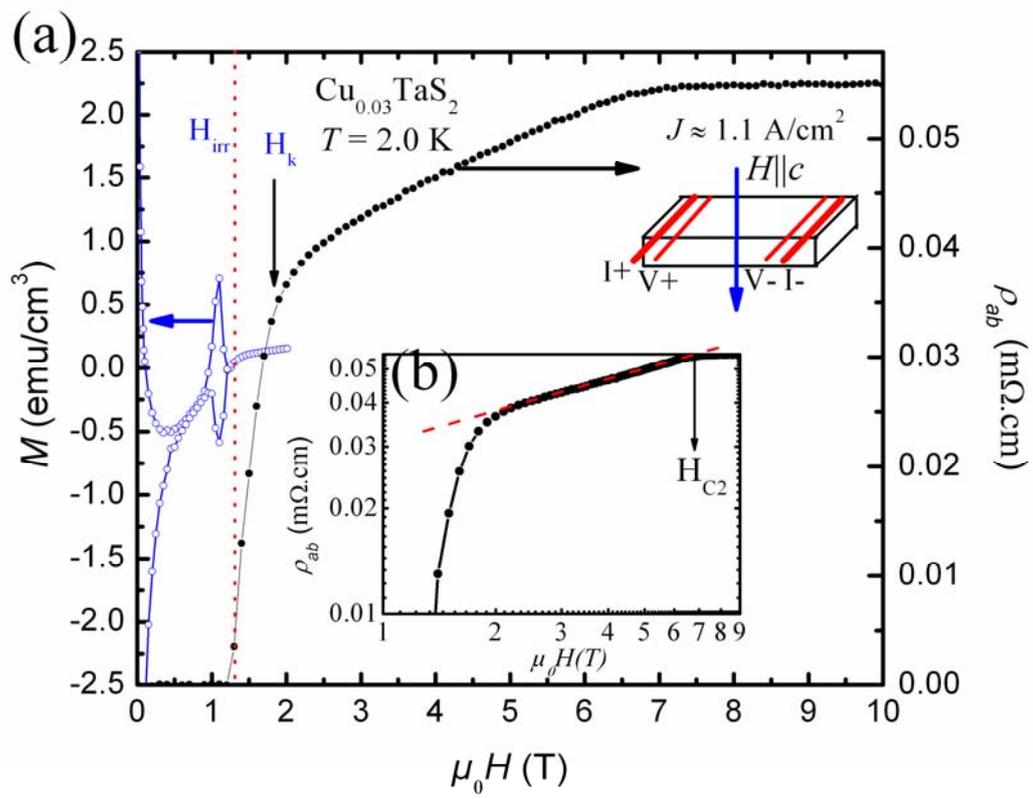

Figure 5 X. D. Zhu *et al.*

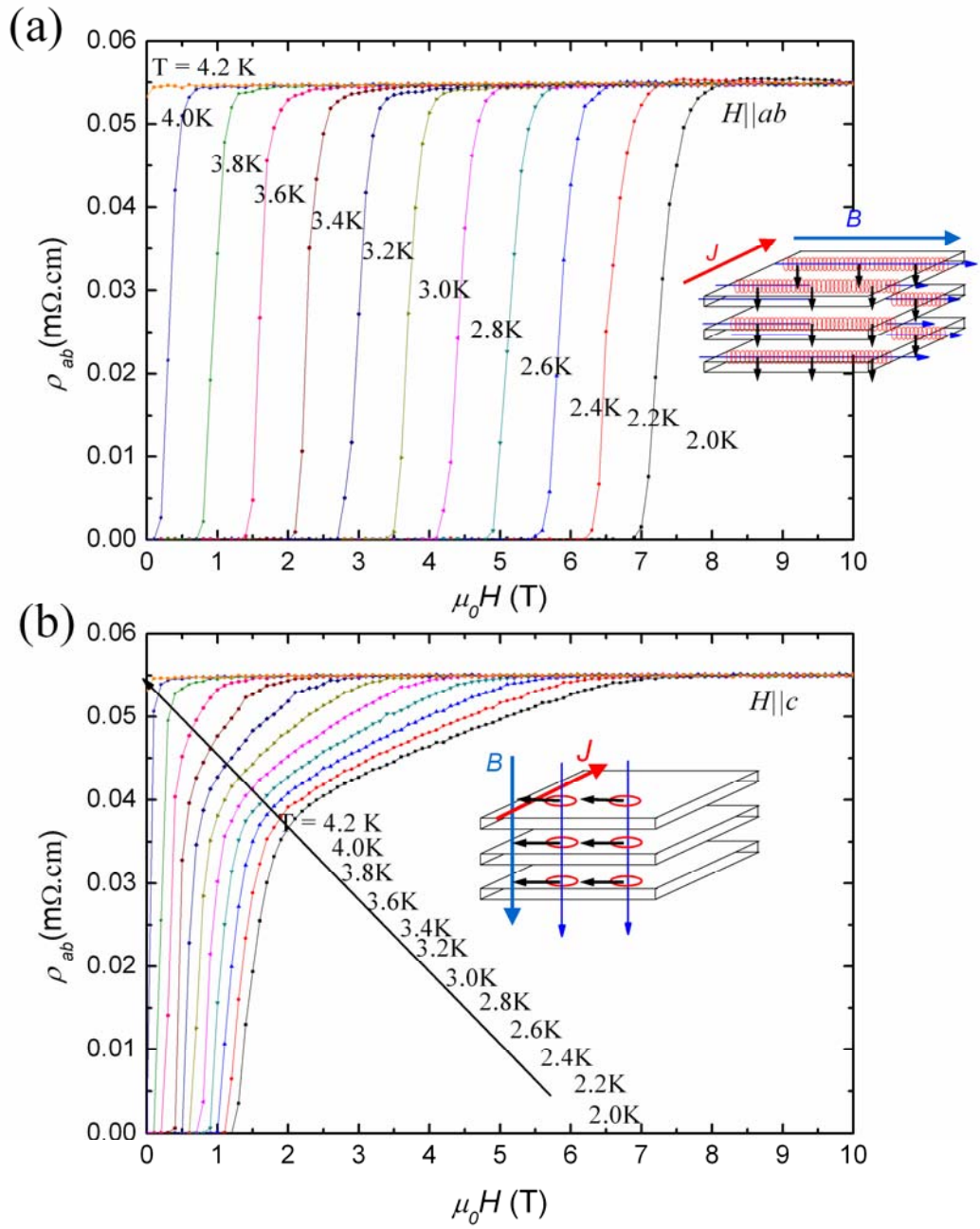

Figure 6 X. D. Zhu *et al.*

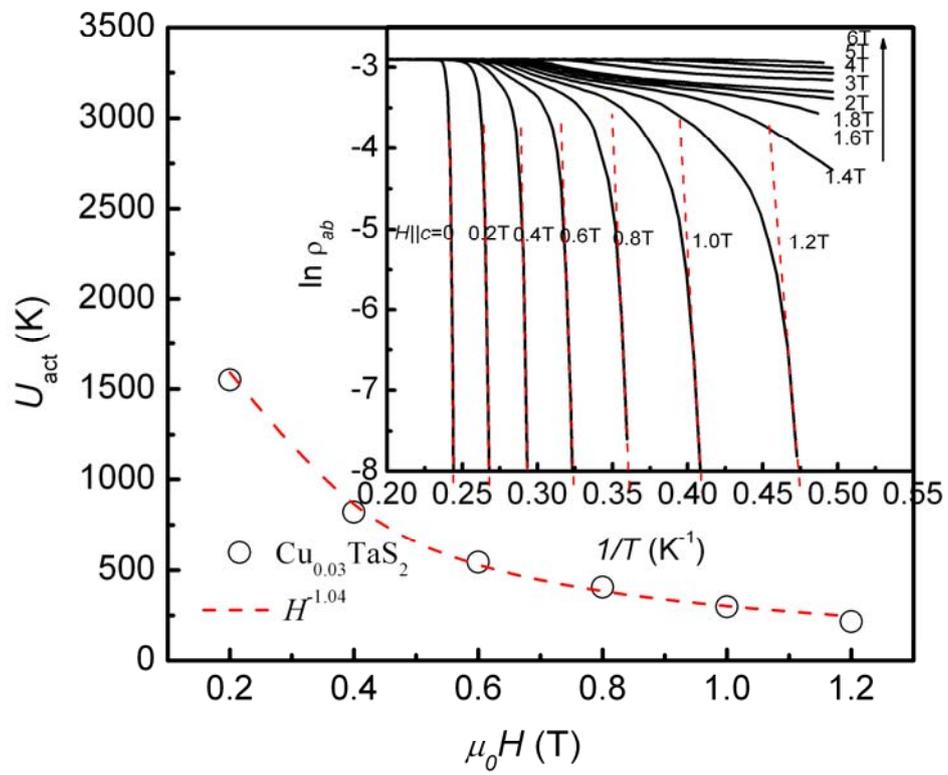

Figure 7 X. D. Zhu *et al.*

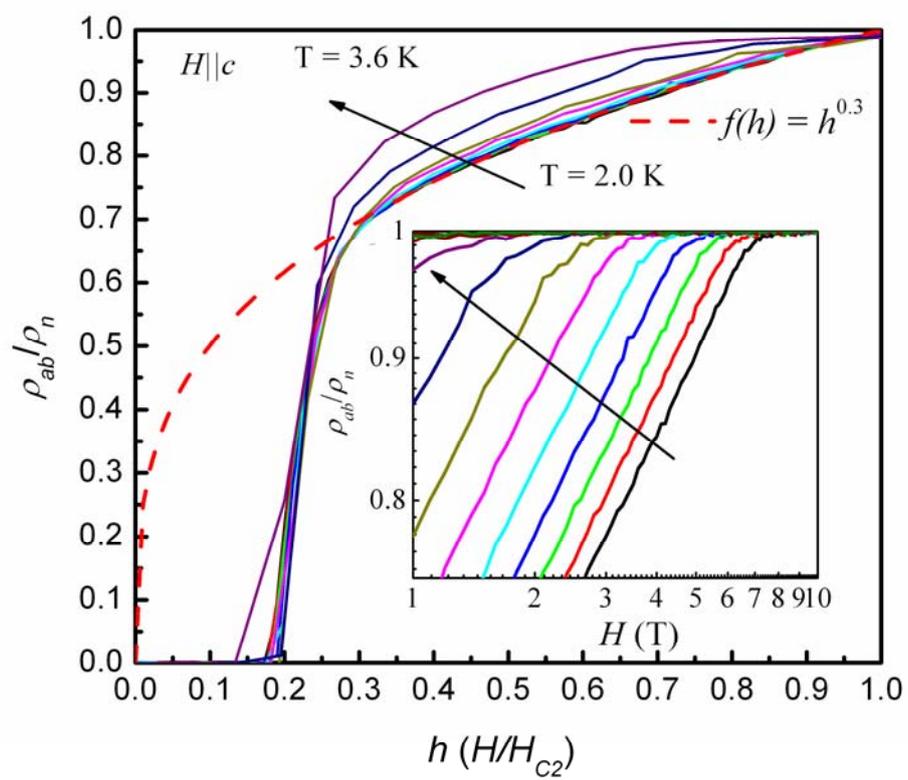

Figure 8 X. D. Zhu *et al.*

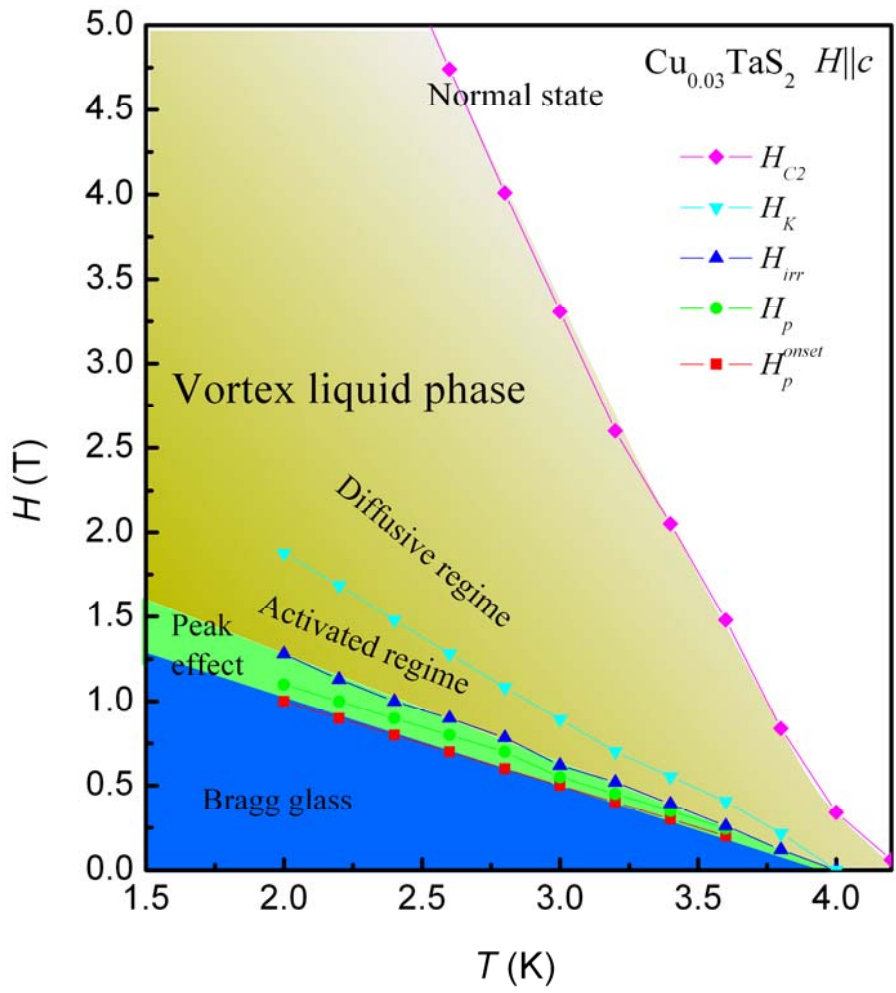

Figure 9 X. D. Zhu *et al.*